\documentclass[%
 reprint,
superscriptaddress,
 amsmath,amssymb,
 aps,prl
]{revtex4-2}

\def\be{\begin{equation}}       \def\ee{\end{equation}}
\def\bea{\begin{eqnarray}}      \def\eea{\end{eqnarray}}
\def\bp{\begin{pmatrix}} \def\ep{\end{pmatrix}}
\def\beaa{\begin{equation}\begin{aligned}}
		\def\eeaa{\end{aligned}\end{equation}}

\usepackage{graphicx}
\usepackage{dcolumn}
\usepackage{bm}
\usepackage{color}
\usepackage[colorlinks,linkcolor=blue,anchorcolor=blue,citecolor=blue,urlcolor=blue]{hyperref}

\begin{document}

\title{Quantum Phonon Dynamics Induced Spontaneous Spin-Orbit Coupling}

\author{Xiangyu Zhang}
\affiliation{New Cornerstone Science Laboratory, Department of Physics, School of Science, Westlake University, Hangzhou 310024, Zhejiang, China}
\author{Da Wang}  
\email{dawang@nju.edu.cn}
\affiliation{National Laboratory of Solid State Microstructures $\&$ School of Physics, Nanjing University, Nanjing 210093, China}
\affiliation{Collaborative Innovation Center of Advanced Microstructures, Nanjing University, Nanjing 210093, China}
\author{Congjun Wu} \email{wucongjun@westlake.edu.cn}
\affiliation{New Cornerstone Science Laboratory, Department of Physics, School of Science, Westlake University, Hangzhou 310024, Zhejiang, China}
\affiliation{
Institute for Theoretical Sciences, Westlake University, Hangzhou 310024, Zhejiang, China}
\affiliation{Key Laboratory for Quantum Materials of Zhejiang Province, School of Science, Westlake University, Hangzhou 310024, Zhejiang, China}
\affiliation{Institute of Natural Sciences, Westlake Institute
for Advanced Study, Hangzhou 310024, Zhejiang, China}

\begin{abstract}
Spin-orbit coupling in solids is typically
a single-body effect arising from relativity.  
In this work, we propose a spontaneous generation of spin-orbit coupling from symmetry breaking.
A spin-dependent electron-phonon coupling model is investigated on a half-filled square lattice, which is solved by sign-problem-free quantum Monte Carlo simulations.
The phase diagram as function of phonon frequency $\omega$ and coupling constant $\lambda$ is fully investigated.
The spin-orbit coupling emerges as an order in the ground state for any $\lambda$ in the adiabatic limit, accompanied by a breathing mode of lattice distortion and a staggered loop spin-current.
This phase dominates in the entire range of $\omega$ with $\lambda< \lambda_{\infty}$, a critical value in the $\omega \to \infty$ limit.
With increasing $\omega$ and $\lambda > \lambda_{\infty}$, the emergent spin-orbit coupling is suppressed and a phase transition occurs leading to charge-density-wave degenerate with superconductivity order.
Our work opens up the possibility of hidden spin-orbit coupling in materials where it is otherwise forbidden by lattice symmetry and paves the way to explore new usable materials or devices in spintronics.
\end{abstract}
\maketitle


\paragraph{Introduction.}
Spin-orbit coupling (SOC), typically originated from relativistic physics \cite{Yang2019}, plays an essential role in spintronics \cite{Hirohata2020,Premasiri2019}.
It has also brought novel quantum states of matter including topological insulators \cite{Kane2005,Kane2005a,Zhang2006} and chiral magnets \cite{Xi2017}. 
In solid systems, the manifestation of SOC usually requires spatial asymmetry.
When a bond does not possess an inversion center, the hopping term across it includes a spin-current term whose spin orientation is determined by the Dzyaloshinskii-Moriya (DM) vector \cite{Dzyaloshinsky1958,Moriya1960}.
Similarly, both the Dresselhaus \cite{Dresselhaus1955} and Rashba \cite{Rashba1984} SOCs also require a noncentrosymmetric crystalline structure.
In these cases, the resultant SOC is a single body effect.
On the other hand, SOC could be generated as an order parameter spontaneously by electron-electron interaction based on Fermi liquid instabilities proposed by one of the authors and his collaborators \cite{congjun2004,Congjun2007}. 
This mechanism also yields novel types of SOCs breaking time-reversal symmetry  beyond the relativistic mechanism exhibiting spin-group symmetries and the ``altermagnetism"-like phenomena later discovered in solid state materials \cite{Libor2022,Mazin2022}.

Electron-phonon coupling (EPI) is an important 
interaction in condensed matter systems, which could induce spontaneous lattice distortions. 
For example, in quasi one-dimensional (1D) systems like polyacetylene, the longitudinal phonons couple to the real part of hopping integrals.
Spontaneous dimerization, {\it i.e.}, the bond-wave instability, occurs as described by the Su-Schrieffer-Heeger (SSH) model \cite{SSH1979}. 
This model has been recently generalized to two dimensions (2D) to study the competition between the bond-wave and other quantum phases such as antiferromagnetism and superconductivity
\cite{Steven2020,Scalettar2021,CaiXun2021,Assaad2022}.
Nevertheless, this kind of electron-phonon interaction is typically spin-independent. 

\begin{figure}[bp]
\includegraphics[width=0.9\linewidth]{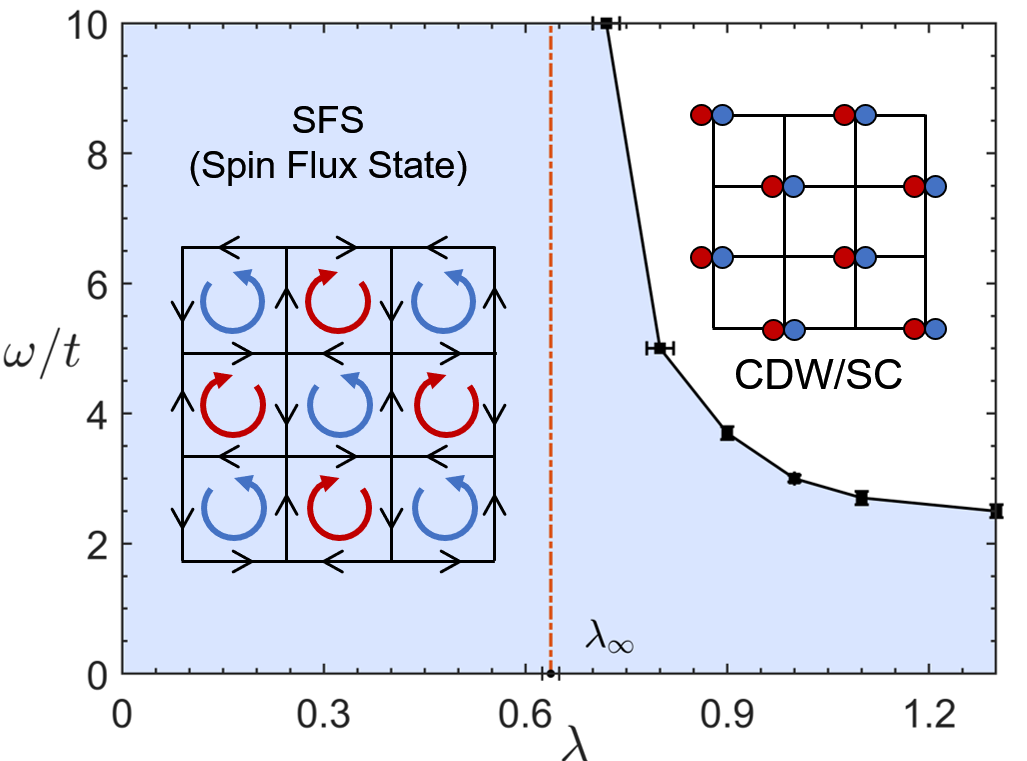}
\caption{\label{fig:pd} 
Phase diagram of the spin-dependent EPC model versus 
coupling constant $\lambda$ and phonon frequency $\omega$.
Red dotted line marks the critical strength $\lambda_\infty$ at $\omega = \infty$.
The two insets illustrate the SFS with loop spin current and degenerate CDW/SC order (represented by the CDW pattern).
 }
\end{figure}

In this article, we generalize the SSH-type Hamiltonian by coupling phonons to the imaginary part of hopping integrals. 
Due to time-reversal symmetry, the coupling form should be like a vibrating spin-current along the bond. 
This type of spin-dependent electron-phonon interaction can induce a new kind of symmetry-breaking phase with a spontaneous generation of SOC. 
We consider the case at half-filling in a square lattice with each bond coupled to an optical phonon mode characterized by the phonon frequency 
$\omega$ and the EPI constant $\lambda$.
The sign-problem free projective quantum Monte Carlo (PQMC) method is employed to yield the ground-state phase diagram shown in Fig.~\ref{fig:pd}. 
In the adiabatic limit, {\it i.e.}, $\omega \to 0$, the system spontaneously develops the spin-flux state (SFS) in which a staggered loop spin current circulating around each plaquette.
This phase has also been  known as spin nematic state\cite{Nersesyan1991,Capponi2004}.
In the anti-adiabatic limit ($\omega \rightarrow \infty$), the SFS still exists below a critical 
EPI constant $\lambda_{\infty}$, above which the system enters the fully gapped phase featuring degenerate 
charge-density-wave (CDW) and superconductivity (SC)
as protected by the pesuod-spin SU(2) symmetry.
Our work here for the first time in a numerical exact way investigates the spontaneous generation of SOC
as the consequence of spin-dependent EPI.

\begin{figure}[tp]
\includegraphics[width=0.8\linewidth]{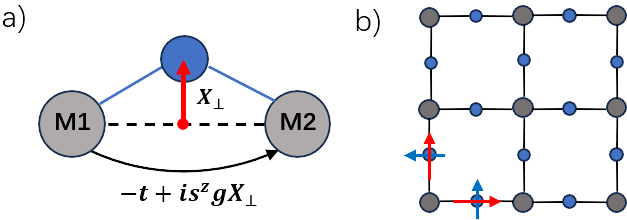}
\caption{\label{fig:2site} (a) A two site case to illustrate how the perpendicular displacement of a middle anion can induce SOC effect, with two transition metal cations M1, M2 (grey) and the anion (blue) between them. 
(b) Square lattice with bond oscillators. Blue and red arrows indicate the positive direction of bond phonon operator $\hat{X}$ and spin current operator $J^{z}$, respectively. }
\end{figure}

\paragraph{Model Hamiltonian}
We first consider a single bond and derive the form of spin-dependent EPI. 
As illustrated in Fig.~\ref{fig:2site} ($a$), when the anion  sits at the bond center of two transition metal cations, due to the inversion symmetry the electron hopping between two cation sites is spin-independent, which reads $H_{el} = -t \sum_{\sigma}(c_{1 \sigma}^{\dagger} c_{2 \sigma}+c_{2 \sigma}^{\dagger} c_{1 \sigma})$, where $t$ is the real part of the electron hopping integral. 
If the anion's displacement is perpendicular to the bond, it breaks the inversion symmetry which results in the DM vector for spin-dependent hoppings. 
Since the residual reflection symmetry with respect to the cation-anion plane, the DM vector is perpendicular to this plane set as the $z$-direction\cite{Moriya1960,Shekhtman1992}.
The resultant spin-dependent EPC term, to the first order of the anion displacement ${X_{\perp}}$, takes the form of 
\begin{eqnarray}
H_{ep}  = g  {X_{\perp}}  {J}_{12}^{z},
\label{eq:Hint1}
\end{eqnarray}
where $J_{12}^z={J}_{12,\uparrow} -  {J}_{12,\downarrow}$ with
${J}_{12,\sigma} = -\mathrm{i}(c^{\dagger}_{1\sigma} c_{2\sigma} - c^{\dagger}_{2\sigma} c_{1\sigma})$;
the coupling strength $g$ is set to be real and positive.

Next we construct a tight-binding model of spin-1/2 electrons on the half-filled $L\times L$ square lattice as sketched in Fig.~\ref{fig:2site}($b$). 
The spin-dependent EPC of Eq. \ref{eq:Hint1} 
and the phonon dynamics on each bond are included.
The total Hamiltonian reads $H = H_{ep}+ H_{ph}$ defined as
\begin{eqnarray}
\begin{aligned}
H_{ep}= &  -t \sum_{\langle ij\rangle, \sigma}\left(c_{i \sigma}^{\dagger} c_{j \sigma}+c_{j \sigma}^{\dagger} c_{i \sigma}\right) +
\sum_{\langle i j \rangle} g\hat{X}_{ij} {J}_{ij}^{z} \\
H_{ph}= &  \sum_{\langle i j\rangle}\frac{\hat{P}_{i j}^2}{2 M}+\frac{1}{2} K \hat{X}_{i j}^2.
\end{aligned}
\label{eq:Hamilton}
\end{eqnarray}
The $H_{ph}$-term describes phonon dynamics
of anions, where $\hat{X}_{ij}$ and $\hat{P}_{ij}$ are conjugate displacement and momentum, $M$ denotes the effective mass and $K$ the effective spring constant. 
The phonon is assumed to be the Einstein-type with a single frequency given by $\omega = \sqrt{K/M}$.
The phonon displacements are restricted to be in-plane for simplicity.
Throughout this work, a dimensionless EPC constant is defined as $\lambda = g/\sqrt{tK}$. 
We set $t = 1$ as energy unit and $K = 1$ by appropriately rescaling $\hat{X}_{i j}$.
The convention of the positive directions 
of phonon displacement and spin current is
illustrated in Fig.~\ref{fig:2site}($b$).

The above spin-dependent EPC describes a dynamic spin-orbit coupling generated by the optical phonons. 
If $\langle {X_{\perp} }\rangle$ vanishes, the SOC coupling 
on average disappears at the single particle level.
On the other hand, if spontaneous 
symmetry breaking occurs with non-vanishing $\langle X_{\perp} \rangle$, SOC will emerge as a consequence of long-range ordering. 
Eq. (\ref{eq:Hamilton}) breaks the spin SU(2) symmetry but maintains the rotational symmetry around the $z$-axis.
Please note that $J_{ij,\sigma}$ should be viewed as a ``canonical" version rather than the physical spin current. 
The physical version $\tilde{J}_{ij,\sigma}$ satisfying the continuity equation is defined in Eq. (\ref{eq:Jc}) later.

The Hamiltonian of Eq. (\ref{eq:Hamilton}) also possesses the pseudo-spin SU(2) symmetry generated by
${\boldsymbol{\eta}} = \frac{1}{2} \sum_{i}{\tilde{c}}^{\dagger}_{i} \boldsymbol{\sigma} {\tilde{c}}_{i}$ with $\tilde{c}^{\dagger}_{i} = (c_{i,\uparrow}^{\dagger}, (-1)^{i}c_{i,\downarrow})$ \cite{CNyang1989}.
The diagonal component of the SU(2) generators
is the total particle number up to a linear relation $\eta^{z} = \frac{1}{2} \sum_{i\sigma}( c^{\dagger}_{i,\sigma} c_{i,\sigma} - 1/2)$, and its off-diagonal components can be reorganized as the $\eta$-pairing operators,
\begin{eqnarray}
\eta^{+} = \sum_{i} (-1)^i c^{\dagger}_{i,\uparrow} c^{\dagger}_{i,\downarrow}, \quad \eta^{-} = (\eta^{+})^{\dagger}
\end{eqnarray}
The pseduo-spin SU(2) unifies the CDW order ${1}/{L^2}\sum_{i,\sigma}(-1)^{i} \langle c^{\dagger}_{i\sigma}c_{i\sigma} \rangle$ and the onsite SC order ${1}/{L^2}\sum_{i} \langle c^{\dagger}_{i\uparrow}c^{\dagger}_{i\downarrow} \rangle$\cite{zhang1990}. 
Moreover, the Hamiltonian is invariant under time reversal $\hat{T} = i\sigma^{y} \hat{K}$ with $\hat{K}$ the complex conjugation.
Due to the Kramers degeneracy, this model is free of sign problem \cite{Congjun2005} and thus can be studied by projective quantum Monte Carlo method with high numerical accuracy. 
\begin{figure}[tp]
\includegraphics[width=\linewidth]{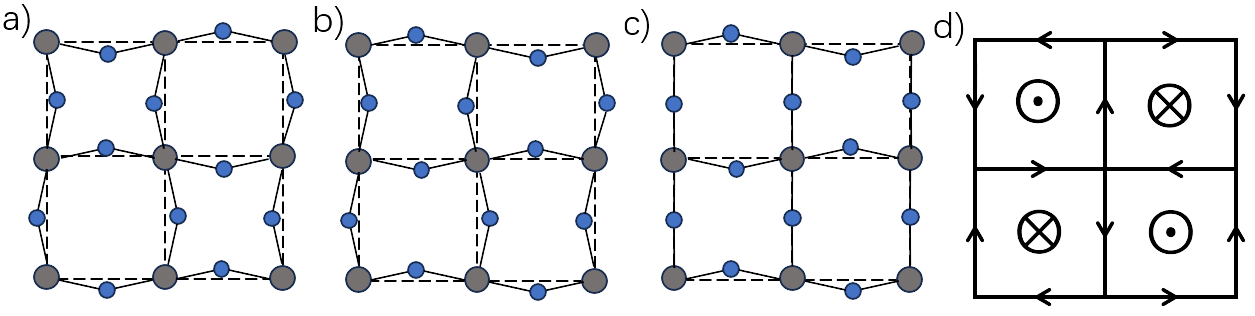}
\caption{Typical lattice distortion patterns $\{X_{ij}\}$ with $\boldsymbol{Q} = (\pi,\pi) $. 
($a$) Breathing mode:  $X_{i,i+\hat{x}} = -X_{i,i+\hat{y}} \neq 0$;  ($b$) Tilting mode: $X_{i,i+\hat{x}} = X_{i,i+\hat{y}}\neq 0$; 
($c$) Unidirectional mode for x-directional bonds only.
($d$) The loop spin-current of SFS state. 
Due to the distortion patterns in ($a$), spin-up and down electrons feel flux in opposite directions around each plaquette, and that for spin-up electrons is marked by $\odot$ or $\otimes$ in ($d$), respectively.
}
\label{fig:configs}
\end{figure}

\paragraph{Adiabatic and anti-adiabatic limits.} 
Phonons with finite frequency mediate a retarded electron-electron interaction.
Before performing QMC simulations, we first investigate the ground state properties in the adiabatic limit, {\it i.e.}, $\omega \rightarrow 0$, where phonon dynamics is classical, and in the anti-adiabatic limit, {\it i.e.}, $\omega \rightarrow \infty$, where the effective interaction between electrons become instantaneous.

In the adiabatic limit, the phonon displacements $X = \{X_{ij}\}$ can be treated as classical variables with static values.
Since the Fermi surface is perfectly nested with the wave vector $\boldsymbol{Q} = (\pi,\pi)$, it is natural to expect that a staggered distortion is most favourable to lower the energy. 
For this nesting $\boldsymbol{Q}$, there are still several possible configurations as shown in Fig.~\ref{fig:configs}($a$), ($b$), ($c$), respectively. 
For all $\lambda$ considered, the breathing mode yields the lowest energy, and results associated with these lattice distortion configurations are shown in Fig.~\ref{fig:MF}($a$).
See Supplementary Material (S. M.) \cite{appdix} for detailed calculations. 

The ground state with the breathing mode configuration is denoted as the spin flux state (SFS).
The lattice distortion induces a staggered SOC with the strength $g|X|$ plotted in Fig.~\ref{fig:MF}($b$).
The effective SOC generates gapless Dirac points at $(\pi/2,\pm \pi/2)$ in momentum space,   which lowers the density of states and saves the energy. 
This SOC pattern corresponds to a staggered spin-flux in each plaquette as illustrated in Fig.~\ref{fig:configs}($d$).
Consequently, the hopping integral is  modified to $t_{ij,\sigma} = -|t_{ij}|e^{\mathrm{i} \theta_{ij,\sigma}}$
where $\theta_{ij,\uparrow} = -\theta_{ij,\downarrow}$. 
Therefore, an electron acquires the spin-dependent phase of
$\Phi_{\sigma} = \sum_{\square} \theta_{ij,\sigma}(\text{mod} \ \pi)$ around a plaquette, where  $\Phi_{\uparrow} = - \Phi_{\downarrow}$.
The time reversal symmetry is preserved since spin up and down currents flow in opposite directions. 
As a comparison, the tilting mode in Fig.~\ref{fig:MF}(b) bears no net flux in any plaquette and thus is equivalent to a trivial case up to a gauge transformation, which, hence, cannot open a gap to lower the electron energy.

In the anti-adiabatic limit, the phonon degrees of freedom can be integrated out to obtain an instantaneous two-body interaction for electrons as
\begin{equation}
 \begin{aligned}
 H_{int}^{eff} & =   - \frac{V}{2}
 \sum_{\langle ij\rangle}(J_{ij,\uparrow}^{z} - J_{ij,\downarrow}^{z})^{2} \\
& = 2V \sum_{\langle ij \rangle}
\left (
\boldsymbol{\eta}_{i} \cdot \boldsymbol{\eta}_{j} + {S}^{z}_{i} {S}^{z}_{j} - {S}^{x}_{i} {S}^{x}_{j} - {S}^{y}_{i} {S}^{y}_{j}
\right ),
     \end{aligned}
\label{eq:Heff}
\end{equation}
where $\boldsymbol{\eta}_i = \frac{1}{2}\tilde{c}^{\dagger}_{i} \boldsymbol{\sigma} \tilde{c}_{i}$ and $\boldsymbol{S}_i  = (S_{i}^x,S_{i}^y,S_{i}^z) = \frac{1}{2}{c}^{\dagger}_{i} \boldsymbol{\sigma} {c}_{i}$ are local pseudospin and spin operators;
$V = {g^2}/{K}$ is the strength of phonon-induced electron-electron interaction, playing the role of antiferromagnetic-like pseudospin exchange interaction.
As discussed in Ref. \cite{zhang1990}, this induces a CDW order degenerate with the onsite SC order. 
By the self-consistent mean-field calculation, the energies versus $\lambda$ for the two types of ground state candidates, SFS and CDW/SC, 
are compared as shown in Fig.~\ref{fig:MF}($c$). 
The crossing at $\lambda_{\infty}^{MF} \approx 1.1$ indicates a first order phase transition from SFS to CDW/SC in the infinite frequency limit. 

\begin{figure}
\includegraphics[width=1\linewidth]{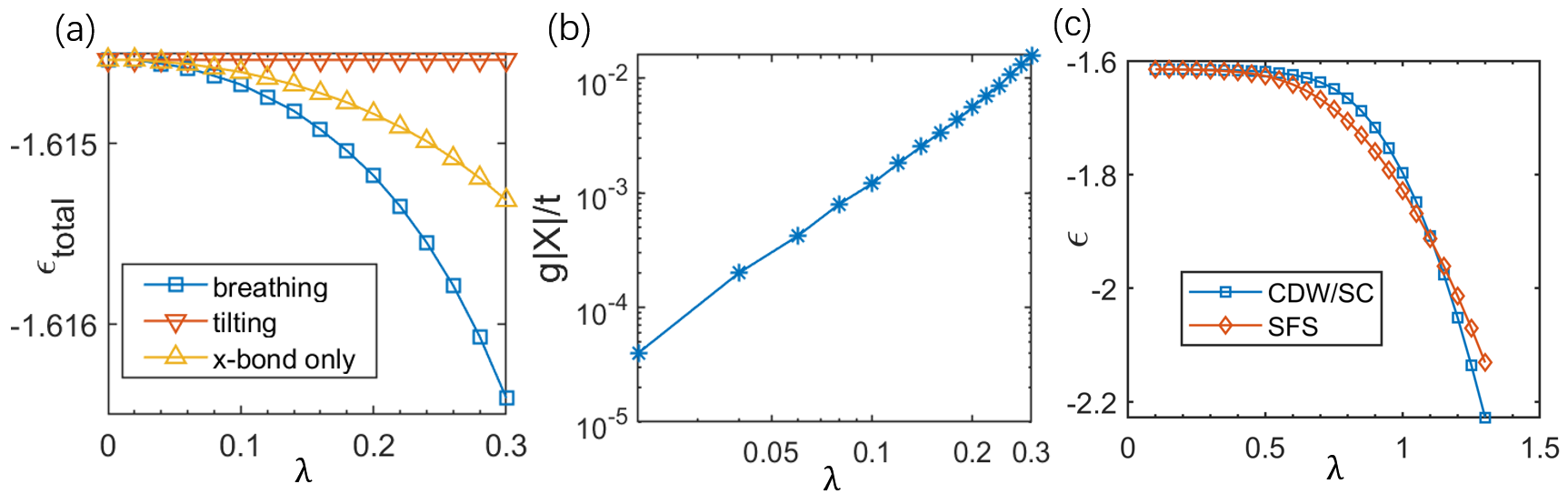}
\caption{Variational calculations 
for $\omega=0$ in ($a$),($b$) and 
the mean field calculations for $\omega=\infty$ in ($c$).
($a$) Energy per site for the three phonon configurations in Fig.~\ref{fig:configs}.
($b$) The strength of emergent SOC $g|X|$ versus EPC constant $\lambda$ in log-log plot.
($c$) Comparison of the mean field energy per site between two ordered phase in the $\omega=\infty$ limit.
}
\label{fig:MF}
\end{figure}

\paragraph{PQMC results.}
For a general phonon frequency $\omega$, we resort to PQMC, which has been shown to yield accurate results for problems of EPC \cite{CaiXun2021,Scalettar2021}.
In order to identify the long-range ordering of the ground state, the structure factor $S(\boldsymbol{Q}) \equiv 1/L^4\sum e^{i \boldsymbol{Q} \cdot \boldsymbol{r}_{ij}}\langle \hat{O}(\boldsymbol{r}_i)\hat{O}(\boldsymbol{r}_j) \rangle$ is calculated for a local observable $\hat{O}(\boldsymbol{r})$.  
The non-vanishing value of $S(\boldsymbol{Q})$ at $L\to\infty$ indicates a long range order.
The structure factors of the bare spin current operator $J^{z}_{x(y)}(\boldsymbol{r})$ and phonon displacement $\hat{X}_{x(y)}(\boldsymbol{r})$ are used to identify the SFS order, where $\hat{O}_{x(y)}$ means the observable is defined along $x(y)$-directional bonds.
In the following, we focus on the ordering momentum $ \boldsymbol{Q} = (\pi,\pi) $ (for SFS or CDW) or $(0,0)$ (for SC).
The phase boundary between SFS and CDW/SC is determined by finite-size scaling extrapolation to the thermodynamic limit. 

\begin{figure} 
\includegraphics[width = \linewidth]{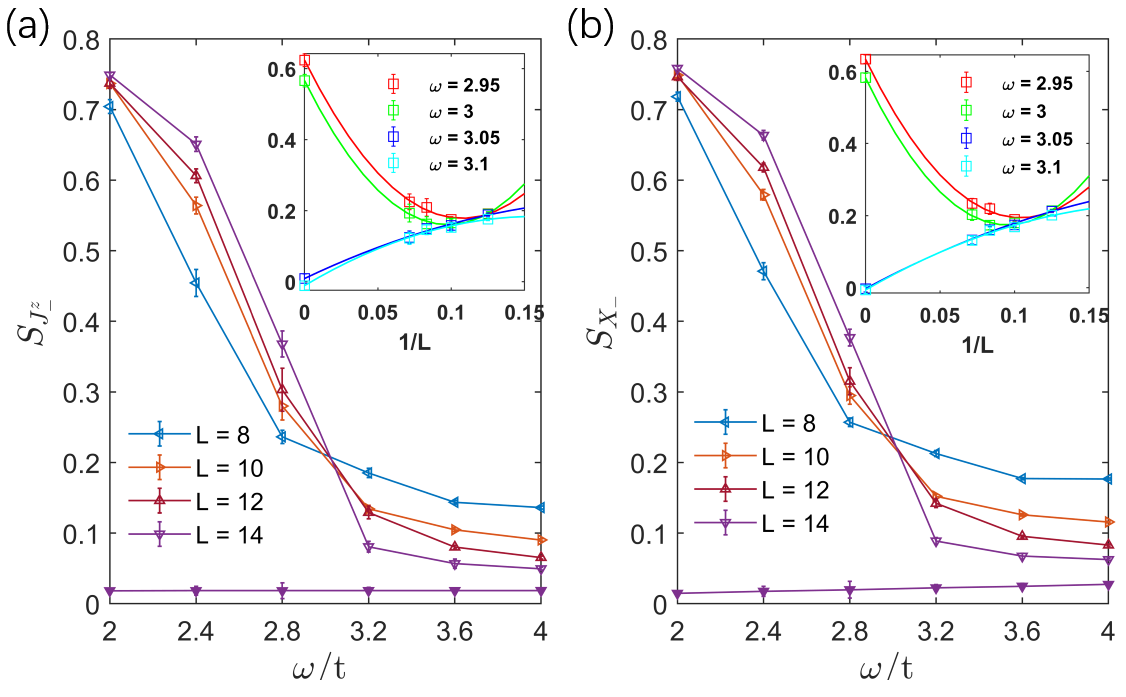}
\caption{Results of the structure factors of $J_{\pm}^{z}$ and $X_{\pm}$ at $\mathbf{Q}=(\pi,\pi)$ with varying $\omega$ and fixed $\lambda = 1$.
Empty (full) markers are used for $S_{J_{-}^{z}} (S_{X_{-}})$ and $S_{J_{+}^{z}} (S_{X_{+}})$. 
Both $S_{J_{-}^{z}}$ and $S_{X_{-}}$ develop a finite value below a critical $\omega_c/t \approx 3.0$, indicating the onset of the SFS order, 
while $S_{J_{+}^{z}}$ and $S_{X_{+}}$ (only values at $L= 14$ are plotted) are nearly zero indicating the negative correlation of observables along $x$- and $y$-directional bonds.
The insets show the extrapolation results close to $\omega_{c}$.
}
\label{fig:S}
\end{figure}

The PQMC simulations are performed for system sizes up to $L=14$.
To reduce finite size effect, all quantities are averaged over twisted boundary conditions (See S. M.\cite{appdix}  ).
The structure factors $S_{J^{z}_{-}}$
and $S_{X_{-}}$ defined for observables
$J^{z}_{\pm} = J_{x}^{z} \pm J_{y}^{z}$ and $X_{\pm} = X_{x} \pm X_{y}$ with varying $\omega$ are presented in  Fig.~\ref{fig:S}($a$) and ($b$), respectively, at the fixed parameter value of $\lambda=1$.
For small values of $\omega$, $S_{J^{z}_{-}}$ extrapolates to finite values, indicating the formation of long range SFS order. 
Instead, for large values of $\omega$, $S_{J^{z}_{-}}$ approaches zero, indicating the vanishing of SFS order. 
The transition occurs at a critical frequency $\omega_{c}/t \approx 3.0$, as shown in the insets of Fig.~\ref{fig:S}.  
The structure factor $S_{X_{-}}$ exhibits similar behaviors to $S_{J_-^z}$, {\it i.e.} yielding the transition frequency $\omega_{c}/t \approx 3.0$ in consistency with that from $S_{J_-^z}$.

In the above finite scaling extrapolations,  the hypothesis of $S(L) \sim a + b/L + c/L^2$ is used. 
As a comparison, the results of $S_{J^{z}_{+}}(S_{X_{+}})$ with system size of $L = 14$ are also shown, which are small for all frequencies and thus rule out both the tilting mode and the unidirectional mode. 
In addition, we have also checked these structure factors at wavevectors of $(0,0)$
and $(\pi, 0)$, and no instabilities are found. 

\begin{figure}[tp]
\includegraphics[scale=0.5]{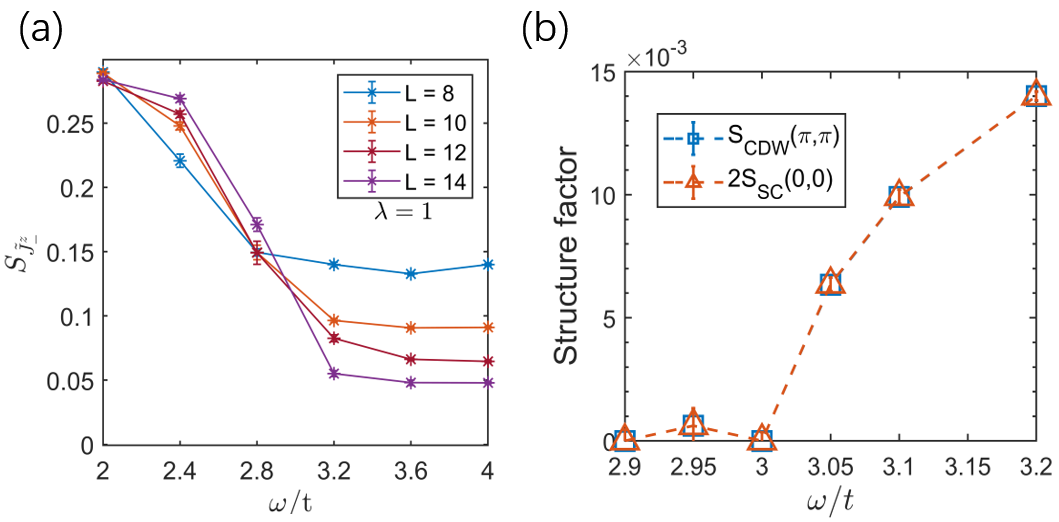}
\caption{($a$) Structure factor at  $\mathbf{Q} = (\pi,\pi)$ of physical version spin-current $\tilde{J}^{z}_{-}$. 
($b$) Extrapolated values of structure factors of CDW/SC.
With $\omega/t > 3.0$, the two orders arise in degeneracy (up to a factor of $2$) guaranteed by the pesudospin SU(2) symmetry. 
}
\label{fig:lambda1pt}
\end{figure}

We examine whether there exists physical circulating spin current around each plaquette. 
Due to $\langle X_{ij} \rangle \neq 0 $, the physical spin current operator on bond $(i,j)$ should be modified to
\begin{eqnarray} 
  \tilde{J}_{ij,\sigma} = -\mathrm{i}(t + \mathrm{i} \nu^{\sigma}g X_{ij})c^{\dagger}_{i,\sigma} c_{j,\sigma}   + h.c.
\label{eq:Jc}
\end{eqnarray}
The result of $S_{{\tilde{J}}^{z}_{-}}(\pi,\pi)$ is plotted in Fig.~\ref{fig:lambda1pt}($a$), showing similar behavior to $S_{J^{z}_{-}}$ in Fig.~\ref{fig:S}($a$).
Although it is suppressed compared to that of $J_{-}^{z}$, they are at the same order, confirming that the SFS phase features a loop spin-current order.

To determine the nature of the phase at $\omega>\omega_c$, we have measured structure factors of multiple possible orders including CDW, onsite SC, antiferromagnetism and ferromagnetism along three spatial directions. 
Only CDW and SC are found exhibiting long range ordering at $\omega > \omega_c$. 
The extrapolated values of structure factors are presented in Fig.~\ref{fig:lambda1pt} ($b$), showing
the degeneracy governed by the pseudospin SU(2) symmetry.


The phase diagram of Fig.~ \ref{fig:pd} based on the PQMC simulations is the main result of this work, in which calculations of  the critical frequency $\omega_c(\lambda)$ are placed together.
Its behavior shows that as increasing $\lambda$ or $\omega$, the SFS phase is suppressed and the CDW/SC order wins. 
Intriguingly, $\omega_c(\lambda)$ diverges as $\lambda$ reduces to a critical value $\lambda_{\infty}$, below which only the SFS order exists in the entire frequency range. 
The value of $\lambda_{\infty}$ is determined by examining the effective Hamiltonian Eq. (\ref{eq:Heff}) in the anti-adiabtic limit, which can also be simulated by the sign-problem free PQMC.
\begin{figure}[bp]
\includegraphics[width = 0.8\linewidth]{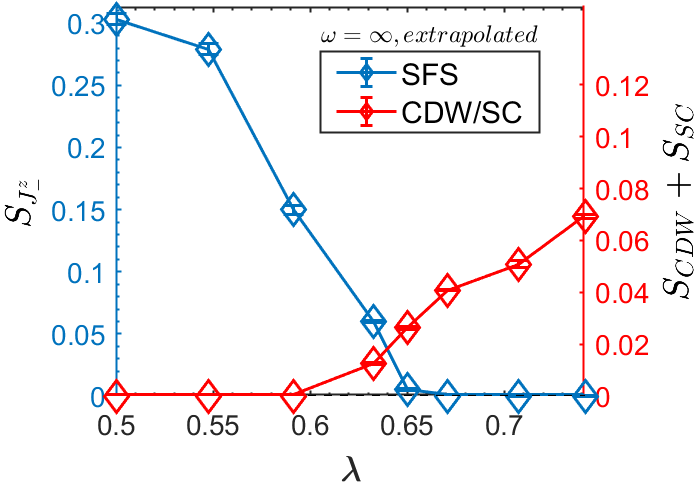}
\caption{ Extrapolation results of SFS and CDW/SC structure factors using data obtained on system size $L = 6 \sim 14$ with $\omega = \infty$.
At $\lambda_{\infty} \approx 0.64$, the crossing of two orders occurs near zero, signifying a phase transition between SFS and CDW/SC orders in the anti-adiabatic limit. 
}
\label{fig:Oinfty}
\end{figure}
The results in Fig.~\ref{fig:Oinfty} show that at $\omega\to \infty$ the transition between these orders occurs at 
$\lambda_{\infty} \approx 0.64$. 
The existence of $\lambda_{\infty}$ also qualitatively agrees with the mean-field calculation for Eq. (\ref{eq:Heff}) with an overestimated value of 
$\lambda_{\infty}^{MF}\approx 1.1$. 

\paragraph{Conclusion and discussion.}
In this article, we have investigated the effect of a spin-dependent electron-phonon interaction on a 2D square lattice at zero temperature by PQMC simulations.
In the SFS phase, a breathing mode of lattice distortions forms due to instability induced by the spin-dependent EPI. 
Consequently a spontaneous SOC effect emerges, giving rise to a staggered spin-opposite flux pattern and circulating spin current in each plaquette.  
It is found that SFS phase persists for all phonon frequencies when $\lambda < \lambda_{\infty}$.
 
Finally, some remarks are given.
It is worth noticing that our theory provides a simplest model to demonstrate a mechanism of phonon-assisted dynamic generation of SOC.
In 1D, the interaction in Eq.(\ref{eq:Hint1}) cannot open a gap thus is not energetically favorable.
In our 2D model, the anion vibrations are constrained within the plane and the motion along z-axis is neglected, which otherwise can induce dynamic DM vector along $x$ and $y$-directions. 
Actually, the staggered displacements of $\rm O^{2-}$ anions along the $z$-axis are observed in the buckled low-temperature orthorhombic (LTO) phase of undoped $\text{La$_{2-x}$Ba$_{x}$CuO$_{4}$}$ \cite{Birgeneau1987,Endoh1988,Congjun2005a}. 
It would be interesting to ask whether the spin-dependent EPC can lead to some observable consequences in this system.

Recently, local inversion-symmetry-breaking lattice distortions has been observed in bulk Bismuthates, such as $\text{Ba$_{1-x}$K$_{x}$BiO$_{3}$}$(BKBO) and $\text{Ba$_{1-x}$Pb$_{x}$BiO$_{3}$}$(BPBO)\cite{Griffitt2023,Edgeton2023}.
Our theory provides an alternative point of view to study the emergence SOC in these materials, which may have potential applications in spintronics.
It was proposed that the so-called hidden SOC could arise in these globally centrosymmetric structures and the large spin-orbit torque with the help of SOC is also detected \cite{Edgeton2023}.
Our study sheds light on the possibility of spin manipulation in centrosymmetric materials, which meets the need of prospering application of spintronic devices.



\textit{Acknowledgements}.
We are grateful to M. Yao, Y. Wang and C. H. Ke for valuable discussions. This work is supported by the National Natural Science Foundation of China under the Grants No. 12234016, No. 12174317.
This work has also been supported by the New Cornerstone Science Foundation.
The computation resource was provided by Westlake HPC Center.


\bibliographystyle{prsty}
\bibliography{epsoc}

\end{document}